\documentclass[jcp, floatfix, nobibnotes, reprint, superscriptaddress]{revtex4-1}
\usepackage{docs}%
\usepackage{siunitx}
\usepackage{amsmath}
\DeclareSIUnit[number-unit-product = {\,}]
\cal{cal}
\DeclareSIUnit\kcal{\kilo\cal}
\DeclareSIUnit[number-unit-product = {\,}]
\Btu{Btu}
\DeclareSIUnit[number-unit-product = {\,}]
\Fahr{\degree F}
\DeclareSIUnit[number-unit-product = {\,}]
\usepackage{bm}%
\usepackage[colorlinks=true,linkcolor=blue]{hyperref}%
\usepackage{graphicx}
\usepackage[utf8]{inputenc}
\usepackage{tabularx} 

\usepackage{dcolumn}
\usepackage{epstopdf}
\usepackage{afterpage}
\usepackage{setspace}
\usepackage{multirow}
\usepackage{dsfont}
\usepackage{sidecap}

\DeclareMathAlphabet\mathbfcal{OMS}{cmsy}{b}{n}
\newcommand\atsign{@}

\setcitestyle{super}

\begin{document}
\title{{\em Ab initio} machine learning of phase space averages}
\author{Jan Weinreich}

\affiliation{University of Vienna, Faculty of Physics, Kolingasse 14-16, AT-1090 Wien, Austria}
\affiliation{University of Vienna, Vienna Doctoral School in Physics, Boltzmanngasse 5, 1090 Vienna, Austria}
\author{Dominik Lemm}

\affiliation{University of Vienna, Faculty of Physics, Kolingasse 14-16, AT-1090 Wien, Austria}
\affiliation{University of Vienna, Vienna Doctoral School in Physics, Boltzmanngasse 5, 1090 Vienna, Austria}
\author{Guido Falk von Rudorff}
\affiliation{University of Vienna, Faculty of Physics, Kolingasse 14-16, AT-1090 Wien, Austria}

\author{O. Anatole von Lilienfeld}
\email{anatole.vonlilienfeld@utoronto.ca}
\affiliation{Machine Learning Group, Technische Universität Berlin, 10587 Berlin, Germany}
\affiliation{Berlin Institute for the Foundations of Learning and Data, 10587 Berlin, Germany.}
\affiliation{Institute of Physical Chemistry and National Center for Computational Design and Discovery of Novel Materials (MARVEL), Department of Chemistry, University of Basel, Klingelbergstrasse 80, CH-4056 Basel, Switzerland}

\date{\today}

\begin{abstract}
Equilibrium structures determine material properties and biochemical functions.
We propose to machine learn phase-space averages, conventionally obtained by {\em ab initio} or force-field based molecular dynamics (MD) or Monte Carlo simulations.
In analogy to {\em ab initio} molecular dynamics (AIMD), our {\em ab initio} machine learning (AIML) model does not require bond topologies and therefore enables a general machine learning pathway to ensemble properties throughout chemical compound space.
We demonstrate AIML for predicting Boltzmann averaged structures after training on hundreds of MD trajectories. 
AIML output is subsequently used to train machine learning models of free energies of solvation using experimental data,
and reaching competitive prediction errors (MAE $\sim$ 0.8 kcal/mol) for out-of-sample molecules-- within milli-seconds.
As such, AIML effectively bypasses the need for MD or MC-based phase space sampling, enabling exploration campaigns throughout CCS at a much-accelerated pace.
We contextualize our findings by comparison to state-of-the-art methods
resulting in a Pareto plot for the free energy of solvation predictions in terms of accuracy and time.
\end{abstract}

\maketitle

\section{Introduction}
\begin{figure*}[htb]
      \centering     
      \includegraphics[width=\linewidth]{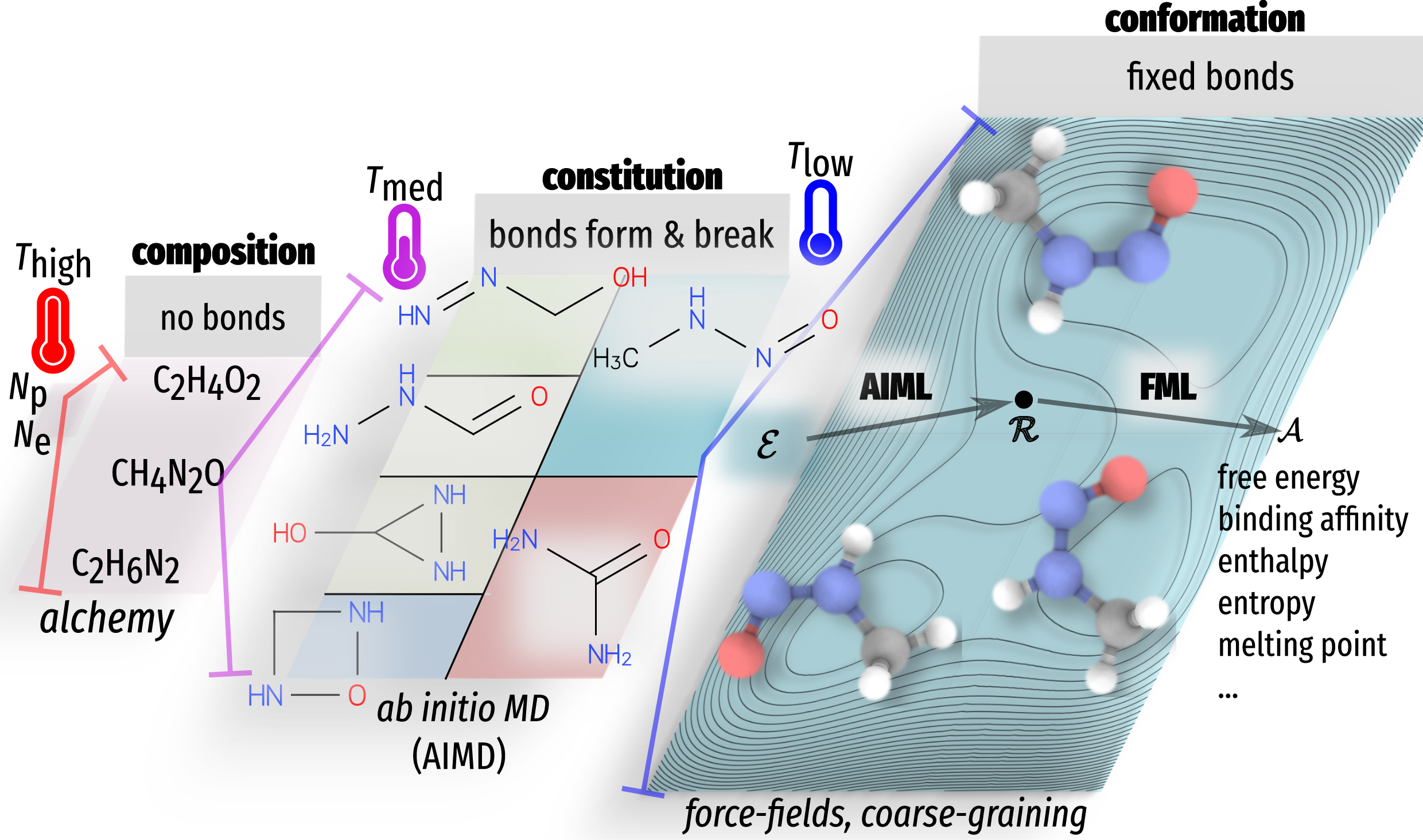}
      \caption{
        Example of a chemical compound space (CCS) for $N_{\rm p} = N_{\rm e} = 32$ protons (and electrons) with a hierarchy of composition, constitution, and conformation. Each level corresponds to distinct temperature-regimes and is described by specific quantum chemical methods\cite{reaxx,car,ostenbrink1,ostenbrink2, amber1, amber2, tip3phase, jorgenson1, jorgenson2,tuckerman}. {\em Ab initio} machine learning (AIML) can act on all three levels and does not require fixed constitutions but allows general ensemble predictions $\mathcal{A}$ using ML-based averaged structures $ \mathbfcal{R}$ and free energy machine learning\cite{fml_paper} (FML).
      }
 \label{fig:fig1} 
\end{figure*}
Structure determines function -- a hallmark paradigm in the atomistic sciences, ranging from biologists studying protein functions based on x-ray structures to organic chemists discussing reaction mechanisms based on NMR measurements\cite{10.5555/1198994}.
The connection between structure $\mathbfcal{R}$ and a compound's function is given through statistical mechanics averages $\mathcal{A}$ over the ensemble $\mathcal{E}$ of Boltzmann weighted configurations,
\begin{align}
   \mathcal{E} \rightarrow \mathbfcal{R} \rightarrow \mathcal{A}~.
    \label{eq:sampling_issue}
\end{align}
The function of a molecule depends on the biological or chemical context e.g. solubilities or binding affinities -- all of which can be expressed as phase space averages $\mathcal{A}$.
Understanding this relation is of fundamental importance as the temperature-dependent balance of configurations dictates the biological function of proteins and their macroscopic behavior (think of egg-white). Unfortunately, to quantitatively predict thermal averages which minimize the free energy imposes major computational challenges due to the necessity of sampling phase space. Furthermore, since experimental efforts to obtain a compounds' structure $\mathbfcal{R}$ are cumbersome several computational routes have been introduced.

However, covering molecular structures poses a monumental challenge. We also highlight the bigger picture of chemical compound space\cite{ccs} (CCS) and its hierarchical structure given by composition, constitution, and conformation. The inherent curse of the dimensions of CCS means that even considering all possible molecules of a single fixed composition quickly results in a combinatorial explosion as illustrated in Fig.~\ref{fig:fig1}. Thus most approaches follow a divide-and--and-conquer strategy addressing the combinatorial problem of CCS at individual levels. Still, the numerical complexity of studying such relationships using molecular dynamics\cite{10.5555/559571} (MD) or Monte Carlo\cite{doi:10.1063/1.1632112,10.1093/biomet/57.1.97} (MC) is overwhelming and to this day most methods with accurate, yet rapid predictions suffer from the curse of conformer sampling. For instance, atomistic simulations study statistical mechanics (SM) ensembles through molecular dynamics $\mathcal{E} \xrightarrow{\text{MD}} \mathbfcal{R} \xrightarrow{\text{SM}} \mathcal{A}$ and are deeply intertwined with insights into biological functions. \textit{Ab initio} molecular dynamics (AIMD) simulations not only allow studying molecules but also chemical reactions\cite{car,PhysRev.136.B864,kresse1,kresse2}. However, they are much more costly
than force fields\cite{polarize, amber1, amber2,doi:10.1021/acs.chemrev.6b00163, reaxx} due to having to solve approximate quantum mechanical equations at every time step. To this account hybrid set-ups $\mathcal{E} \xrightarrow{\text{MD}} \mathbfcal{R} \xrightarrow{\text{ML}} \mathcal{A}$ using both atomistic simulation and machine learning (ML) have been introduced uniting quantum mechanical equations with surrogate \textit{learning on the fly} potentials\cite{LOTF,Zhenwei2015,jinnouchi2019fly}. This helps mitigate some of the \textit{ab initio} costs but may still require extensive MD sampling. These challenges have driven technological advancements of dedicated computer hardware, e.g. of the supercomputer Anton\cite{anton}, specifically designed to accelerate MD simulations. Conversely, large MD codes\cite{amber1,openmm} have been rewritten in CUDA\cite{vingelmann2020cuda} just so that they can run on GPUs. Decentralized global computing network initiatives such as Folding$\atsign$home\cite{folding1, folding2} also predominantly run MD. MD also routinely consumes major fractions of resources and energy costs of high-performance computing centers, as recently seen for the Gordon Bell award to Car and co-workers for running MD on 100 M atoms\footnote{\url{https://www.acm.org/media-center/2020/november/gordon-bell-prize-2020}}.

\begin{figure*}[htb]
          \centering      
          \includegraphics[width=\linewidth]{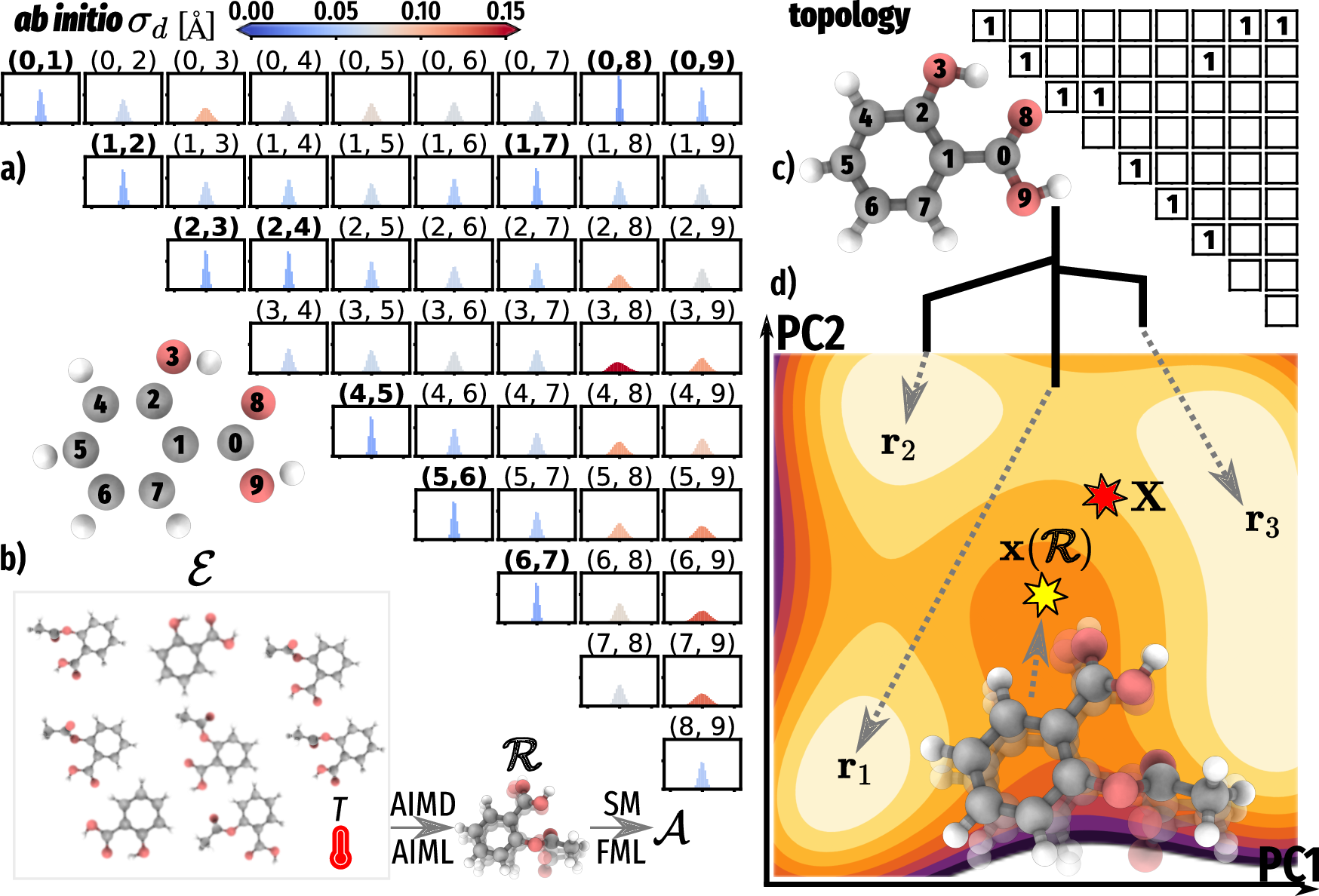}
          \caption{
                Centered histograms and standard deviation $\sigma_d$ of all distances between ten non-hydrogen atoms of aspirin extracted from an \textit{ab initio} molecular dynamics (AIMD) trajectory\cite{mddata} (a).
                Conventional AIMD and \textit{ab initio} ML (AIML) map the ensemble $\mathcal{E}$ to averaged structure $\mathbfcal{R}$ to the statistical mechanics (SM) average $\mathcal{A}$ (b). Bold labeled index pairs \textbf{(i,j)} define a bond topology (c). Sketch of two principal components $\text{PC}1, \text{PC}2 $ of ML-based representations\cite{Parsaeifard_2021,PhysRevB.87.184115, nigam2022unified, representations} on a fictitious free energy surface (d) corresponding to conformers represented by a disconnectivity graph\cite{wales}. Instead of MD average representations\cite{fml_paper} (FML) $\mathbf{X}$ we propose AIML predicted representations of averaged conformers $\mathbf{x} (\mathbfcal{R})$.
          }
     \label{fig:conformers} 
\end{figure*}	
To address the length and timescale problem of conformational space across CCS, we have recently introduced Free energy Machine Learning (FML) which relies on the averaged structure $\mathbfcal{R}$ as input to predict ensemble averages such as the free energy~\cite{fml_paper}, $\mathbfcal{R} \xrightarrow{\text{FML}} \mathcal{A}$. 
Averaging the structure is necessary because ensemble properties inherently depend on multiple configurations and as such using a single geometry per molecule introduces ambiguities to the ML model.
Here, we propose to replace the preceding step, i.e. the generation of the averaged input through extensive molecular dynamics runs for any query compound by an {\em ab initio} machine learning (AIML) model, $\mathcal{E} \xrightarrow{\text{AIML}} \mathbfcal{R}$.

AIML makes use of the Graph-To-Structure\cite{lemm2021machine} (G2S) method to predict three-dimensional structures though chemical compound space. Training data and labels, however, are fundamentally different from G2S: Instead of using a single optimized conformer per molecule, AIML training data consists of averages over complete MD trajectories, enabling the prediction of thermodynamically averaged conformers. This accounts for the fundamentally important difference between a Boltzmann average and a single atomic configuration for ensemble property predictions, as previously discussed in free energy machine learning\cite{fml_paper} (FML).

In particular, we replace ensemble sampling with machine learning of ensemble averages $\mathcal{A}$ of an equilibrium property $a(\mathbf{r}, \mathbf{p})$ by,
\begin{align}
          \mathbfcal{A} =  \; \frac{1}{Z} \int a(\mathbf{r}, \mathbf{p}) ~ e^{-\beta E(\mathbf{r}, \mathbf{p})}~  \text{d} \mathbf{r}~ \text{d} \mathbf{p}  \approx \mathbfcal{A}^{\rm ML} (\mathbfcal{R})\label{eq:ensemble}~,
\end{align}
with $\beta$, $Z$, and $E$ being the Boltzmann-factor, the partition function, and the total energy, respectively\cite{tuckerman2}. The approximate equality of Eq.~\ref{eq:ensemble} is achieved by training an ML model $\mathbfcal{A}^{\text{ML}}$ that uses only the averaged conformer $\mathbfcal{R}$ with values $\mathbfcal{A}$ that include the rigorous integral over the ensemble.

AIML is a purely ML-based framework and properly accounts for the underlying Boltzmann statistics and can subsequently be used to generate the appropriate input for FML model-based predictions. The goal of our work is not to build an ML model that perfectly reflects thermodynamic expectation values but to construct a surrogate model that can predict these integrals with high accuracy and speed. By including Boltzmann averaged conformers we make sure to include a canonical mapping of the underlying ensemble of each molecule to the ensemble property. By training a second FML model on experimental values we ensure that our mapping $\mathbfcal{A}^{\text{ML}}(\mathbfcal{R})$ from the averaged structure also includes contributions from the complete ensemble as defined in the proper phase space integral $\mathbfcal{A}$. Our numerical results, i.e. the systematic improvement of the models' accuracy with training set size, indicate that our assumption (Eq.~\ref{eq:mapping}) is sufficiently valid for the dataset we studied.

After briefly introducing AIML in the following, we will demonstrate its applicability
for the prediction of aqueous free energies of solvation of out-of-sample molecules
{\em without} having to perform explicit MD simulations. 
For training, however, extensive MD trajectories at corresponding temperatures are necessary, as well as experimental measurements of solvation reference energies. 
Lastly, we provide an overview of the efficiency of AIML in the context of alternative 
state-of-the-art solvation methods. We find that AIML offers respective speed-ups by four to seven orders of magnitude when compared to classical or \textit{ab initio} MD-based predictions of free energies of solvation.

\section{Methods}

\subsection{FML and G2S}
\label{sec:g2sANDfml}
    For the previously published free energy machine learning\cite{fml_paper} (FML) approach first all sampled geometries $\mathbf{r}$ had to be transformed to representation vectors $\mathbf{x}(\mathbf{r})$ before finally the average $\mathbf{X}$ over the representation vectors,
\begin{align}
    \mathbf{X} = \int \mathbf{x}(\mathbf{r})~ e^{-\beta E(\mathbf{r})}~\text{d}\mathbf{r}~,
\end{align}
could be computed. The key advantage of AIML is that instead of explicitly sampling conformer space, a single AIML evaluation is required to predict a surrogate vector $\mathbf{x}(\mathbfcal{R})$ evaluated for the system average $\mathbfcal{R}$ to replace $\mathbf{X}$ (s. Fig.~\ref{fig:conformers}d).

Graph-To-Structure (G2S) exploits implicit correlations among relaxed structures in training data sets to infer interatomic distances for out-of-sample compounds across chemical space. G2S effectively enables direct reconstruction of three-dimensional coordinates, thereby allows circumventing conventional energy optimization. G2S can reach an accuracy on par or better than conventional structure generators. As query input, G2S requires only bond-network and stoichiometry-based information. G2S learns the direct mapping from a chemical graph to that structure that had been recorded in the training data set. For the prediction of new structures, only molecular connectivity is needed, which can be provided e.g. via SMILES\cite{doi:10.1021/ci00057a005} or SELFIES\cite{selfie}. The G2S machines predict all pairwise distances. The full 3D geometry is then reconstructed using DGSOL\cite{dgsol} for heavy atoms and a Lebedev sphere optimization scheme for hydrogen atoms.

\subsection{Workflow}
\label{sec:method_g2fml}

 \begin{figure}[htb]
          \centering         
          \includegraphics[width=\columnwidth]{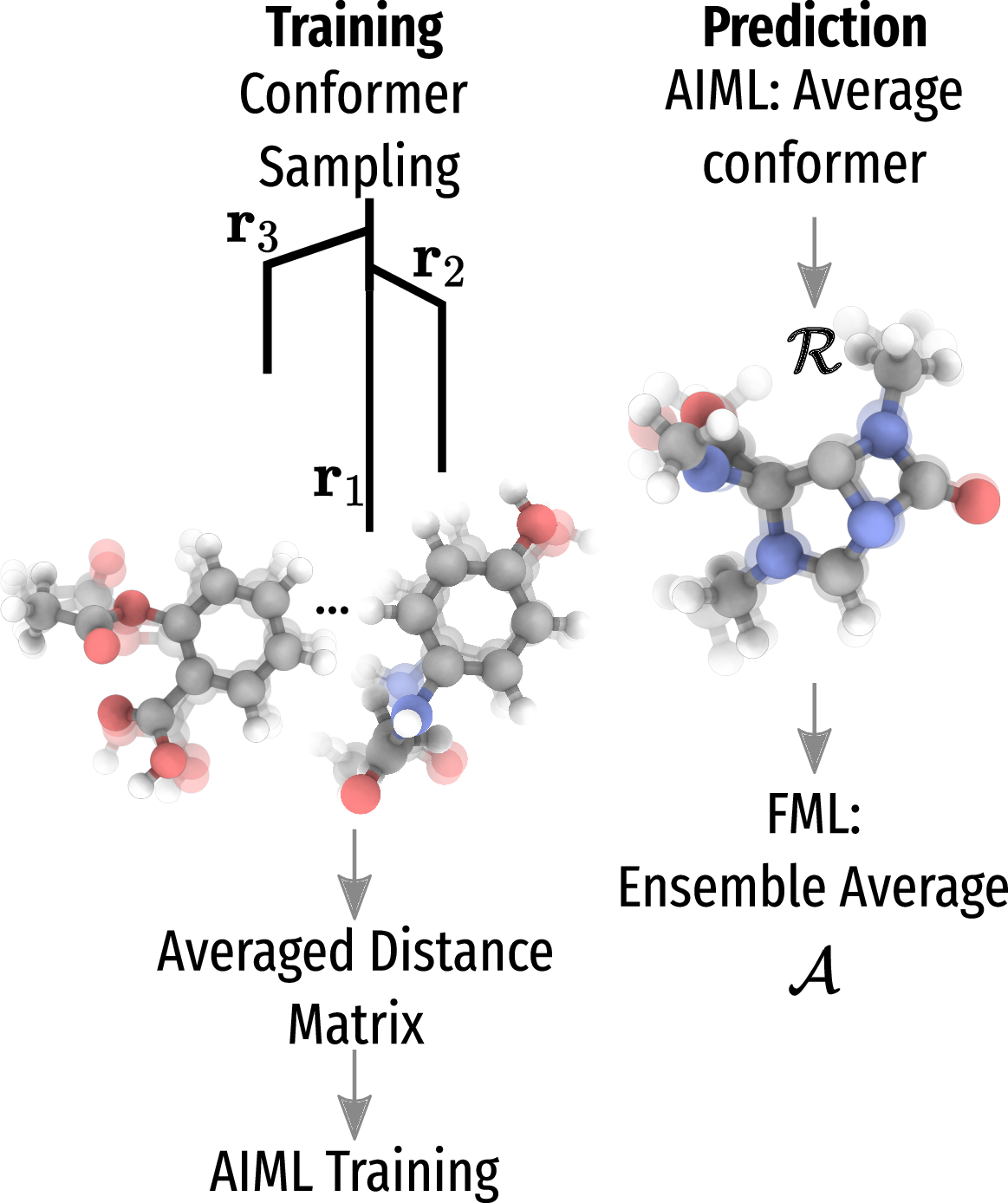}
          \caption{
          To generate {\em ab initio} machine learning (AIML) training data, conformer space is sampled to obtain Boltzmann weighted average distance matrices for given molecular graphs. AIML predicts the averaged distance matrix, which is then converted to a three-dimensional geometry $\mathbfcal{R}$. Finally, the ensemble average $\mathcal{A}$ is predicted using free energy machine learning\cite{fml_paper} (FML).
          }
     \label{fig:g2fml_idea} 
 \end{figure}

An essential ingredient for AIML was extending the previous Graph-To-Structure\cite{lemm2021machine} (G2S) method via the introduction of an averaged geometry $\mathbfcal{R}$. This enables a computationally efficient ML-based map between ensemble and free energies and addresses the conformer sampling bottleneck. AIML provides an effective representation for the conformer ensemble by mapping all degrees of freedom to a single averaged structure to approximate the ensemble-averaged structure with corresponding AIML representation vector $\mathbf{x}$ (cf. Fig.~\ref{fig:conformers}). A schematic overview of the steps for AIML training and prediction is given in Fig.~\ref{fig:g2fml_idea}. As we will discuss in the following two steps are needed to combine both methods: i) use Boltzmann weighted distance matrices $\mathbf{D}$ for training ii) use AIML average conformer predictions as input for an ensemble average model (s.~Fig.~\ref{fig:g2fml_idea} right). The required training steps for the AIML structure prediction use Boltzmann-averaged intramolecular distance matrices. More specifically, as a first step, the molecule is transformed to a graph-based representation\cite{lemm2021machine} $\mathbf{g}$ with the average distance matrix $\mathbf{D}$ as training labels resulting in ML models for heavy atom pairs as well as heavy and hydrogen atoms. Before entering the next step of free energy prediction, AIML is trained with the maximal number of molecules ($N=512$) to construct the average training and test set conformers. This process is repeated for the complete dataset with consistent training test splits between AIML structure and free energies prediction. Next, a machine (s. sec. \ref{sec:krr}) for learning free energies is trained using AIML predicted average geometries. Finally, AIML can be used for out-of-sample predictions (s. Fig.~\ref{fig:g2fml_idea}).
Based on the molecular graph, the Boltzmann weighted distance matrix $\mathbf{D}$ is predicted. Next, a distance-geometry solver\cite{dgsol} (DGSOL) is used to convert the distance matrix to three-dimensional coordinates. The predicted average conformer then serves as a link between the graph and corresponding three-dimensional geometry (s. Fig.~\ref{fig:conformers}). Subsequently, the average predicted conformer $\mathbfcal{R}$ is transformed to a single Bag-of-Bonds\cite{bob} (BoB) representation vector $\mathbf{x}( \mathbfcal{R})$ and used to predict the free energy.

\subsection{Machine learning}
\label{sec:krr} 

ML, in particular of solutions to the Schrodinger equation i.e. quantum machine learning\cite{huang2020ab, doi:10.1063/5.0051418, doi:10.1063/5.0047760, tkatchenko2020machine, doi:10.1021/acs.chemrev.1c00107} (QML), allows navigating chemical compound space (CCS) with high efficiency and precision. ML has become a popular avenue to material science with applications to atomization energies\cite{coulomb, doi:10.1021/acs.jctc.7b00577}, crystal formation energies\cite{PhysRevLett.117.135502}, carbenes\cite{schwilk2020large}, excited states\cite{westermayr2020machine,dral2021molecular}, oxidation states\cite{doi:10.1063/5.0021452}, nuclear magnetic resonance spectra\cite{rhagu}, reaction barriers\cite{vonrudorff2020thousands, bragato2020data}, magnetic systems\cite{eckhoff2021highdimensional} and charge transfer\cite{doi:10.1021/acs.accounts.0c00689} or molecular fragments\cite{superbing, huang2020dictionary}. Many recent approaches are based on individual levels of CCS, i.e. composition\cite{goodall, goodall2021rapid}, constitution\cite{WIEDER2020, PhysRevLett.108.058301, ecfp, doi:10.1063/1.5019779} (s.~Fig.~\ref{fig:fig1}). Moreover ensemble properties such as the free energy of solvation\cite{fml_paper,mlpcm,riniker2017molecular, axelrod2020molecular, doi:10.1021/acs.jcim.0c00479, hybridML, lim2020mlsolva,axelrod2020molecular,vermeire2020transfer}, melting points\cite{jinnouchi2019fly, venkatraman2018predicting, opera}, magnetic anisotropy tensors\cite{doi:10.1021/acs.jctc.1c00853}, phases of water\cite{reinhardt2021quantum, cheng2021predicting,monserrat2020extracting} have previously been addressed with ML.

Our ML approach is based on Kernel-Ridge Regression\cite{Vapnik1998} (KRR) a supervised learning method that allows approximating arbitrary functional relationships between input data given as molecular representations $\mathbf{x}$ and properties $\mathcal{A}(\mathbf{x})$. Using KRR $\mathbf{x}$ is mapped into a high dimensional feature space rendering the regression problem linear. A remarkable result of KRR\cite{Vapnik1998, Deisenroth2020, anatolebook} is that the mapping does not need to be carried out explicitly, instead, the distances between representations $\mathbf{x}_i$ and $\mathbf{x}_j$ are computed e.g. using Gaussian kernel functions,
\begin{align}
    k(\mathbf{x}_i, \mathbf{x}_j) = \exp{\left( -\frac{\vert \vert \mathbf{x}_i - \mathbf{x}_j \vert \vert^{2}_{2}}{2 \sigma^2} \right)}~,
    \label{eq:kernel}
\end{align}
that measure the similarity between two compounds $i$ and $j$ resulting in the kernel matrix $\mathbf{K}$ where $\sigma$ is the kernel-width hyperparameter. We use KRR to predict the vector of all interatomic distances $\mathbf{D}_{q}$ of a query compound $q$ and a distance geometry solver\cite{dgsol} (DSGOL) for subsequent reconstruction of the three-dimensional geometry.

The vector $\mathbf{D}_{q}$ contains all distances of atoms in the query molecule. The average conformer prediction $\mathbfcal{R}(\mathbf{x}_{q} =  \mathbf{g}_{q})$ of a query molecule $q$ represented by a graph\cite{lemm2021machine} $\mathbf{g}_{q}$ is given by,
\begin{align}
     \mathbfcal{R}(\mathbf{g}_{q}) =\text{DGSOL} (\mathbf{D}(\mathbf{g}_{q}))~,
    \label{eq:mapping}
\end{align}
where the distance prediction is as follows,
\begin{align}
         \mathbf{D}_{q} =\mathbf{K} \cdot \boldsymbol{\alpha}~.
\end{align}
Here, the kernel matrix $\mathbf{K}$ is evaluated between query and training compounds $\mathbf{g}_{i}$ and $\mathbf{g}_{q}$ with regression coefficients $\boldsymbol{\alpha}$. The optimal regression coefficients $\boldsymbol{\alpha}$ are obtained by solving a set of equations,
\begin{align}
    \boldsymbol{\alpha} =  (\mathbf{K} + \lambda \cdot \mathds{I})^{-1} \mathbf{D}~,
    \label{eq:solution}
\end{align}
where the vector $\mathbf{D}$ contains all distances between atoms for each of the training molecules and $\lambda$ is a regularization parameter. The ensemble property prediction is given by:
\begin{align}
             \mathbfcal{A}_{q} =
             \widetilde{\mathbf{K}} \cdot \widetilde{\boldsymbol{\alpha}}~.
\end{align}
As before, for training $\widetilde{\mathbf{K}}$ is evaluated between all training compounds now using molecular representation vectors $\mathbf{x}(\mathbfcal{R})$:
\begin{align}
    \widetilde{\boldsymbol{\alpha}} = (\tilde{\mathbf{K}} + \lambda \cdot \mathds{I})^{-1} \mathbfcal{A}~,
\end{align}
where the vector $\mathbfcal{A}$ contains the values of the ensemble property in the training set. Learning curves quantify the model prediction error, often measured as mean absolute error (MAE), against the number of training samples $N$ and are key to understand the efficiency of ML models. It is generally found\cite{Vapnik1998} that they are linear on a log-log scale,
\begin{align}
    \log{ \left( \frac{ \text{MAE} }{\text{unit}} \right) }   \approx I -  S \cdot \log{(N)}~,
\end{align}
where $I$ is the initial error and $S$ is the slope indicating model improvement given more training data.

\section{Results}
\label{sec:results}
\subsection{Concept of equilibrium structure prediction}
\label{sec:structure}
Within AIML, we view the averaged structure as an ensemble property, $\mathbfcal{R} =\mathcal{A}$, representing the connection to the aforementioned overarching theme of structure determining function\cite{10.5555/1198994}. 
In addition, the thermal equilibrium structure is relevant for NMR spectroscopy which accounts for protein flexibility by resulting in time-averaged structures equivalent to $\mathbfcal{R}$ due to the ergodic theorem\cite{tuckerman2}. Purely ML-based implementation bypasses routinely encountered sampling issues -- \textit{de facto} replacing MD simulations with predicted averages as ensemble fingerprints for subsequent property prediction.

We first use kernel-ridge regression~\cite{Vapnik1998} (KRR) to predict the symmetric matrix of averaged interatomic distances of a query compound $q$, i.e.~$\mathbf{D}_{q} \approx \mathbf{K} \cdot \boldsymbol{\alpha}$ which contains all inferred averaged distances of atoms in the query molecule.
$\mathbf{K}$ and $\boldsymbol{\alpha}$ correspond to the kernel matrix and training weights obtained for a training set consisting of MD trajectories and averaged interatomic distances as labels. In analogy to Graph-To-Structure~\cite{lemm2021machine} (G2S), we subsequently rely on the distance geometry solver\cite{dgsol} (DSGOL) for reconstruction of the three-dimensional structure $\mathbfcal{R}$,
as well as on graph-based representations, $\mathbf{g}_q$.
To exemplify the AIML approach, consider the \textit{ab initio} molecular dynamics (AIMD) trajectory published in Ref.~\cite{mddata} of the aspirin molecule at 300 K, resulting atomic distance histograms in Fig.~\ref{fig:conformers}a). 
In order to establish a graph-based representation to replace \textit{ab initio} MD with \textit{ab initio} ML (AIML) as illustrated in Fig.~\ref{fig:conformers}b, it is necessary to assign bonds. This is straightforward using the distance histogram as covalently bonded atoms will not move far from each other (see Fig.~\ref{fig:conformers}c)) -- a concept that can be generalized via coarse-graining\cite{doi:10.1021/acs.chemrev.6b00163, kolinski2004protein,liwo2014unified, levitt1975computer}. While AIML requires a suggested molecular graph it is not restricted to a single fixed bond topology but allows adapting the molecular graph depending on the relevant degrees of freedom depending on temperature or the environment of the molecule. Thus AIML can include the formation and breaking of bonds (cfg. Fig.~\ref{fig:fig1}) corresponding to adding or erasing a one in the bond topology matrix $\mathbf{g}_q$. (cfg. Fig.~\ref{fig:conformers}). Note that the predicted pairwise distance matrix does not only account for nearest neighbor effects but the complete many-body description since it includes all cross combinations of atomic distances.

Using the graph as the representation for constructing kernels, the AIML model then learns the center of each off-diagonal element in the distance histogram as a label.
Next, the AIML predicted distances are used to reconstruct the average conformer. Subsequently the ML representation vector $\mathbf{x}(\mathbfcal{R})$ is computed. Therefore, AIML allows exchanging the order of average evaluation compared to the previous FML\cite{fml_paper} approach (illustrated Fig.~\ref{fig:conformers}) resulting in a dramatic reduction of computational costs.

AIML proposes a different paradigm by connecting all hierarchies of CCS with ensemble properties into a single ML-based framework. Because of its generality, AIML does not require \textit{a priori} information about bonds but only averaged atomic distances. In analogy to the hierarchy of CCS, AIML has several special cases with fundamental physical interpretations (s. Fig.~\ref{fig:fig1}): At temperatures higher than most bond energies AIML operates on atomic clusters corresponding to a topology matrix that (mostly) contains zeros i.e. the composition. At moderate temperatures, bonds exist but may occasionally break corresponding to adding or removing a zero in the topology matrix. In this case, multiple molecular graphs can be extracted and AIML predicts the constitution averaged structure.

\subsection{Application: Ensemble to structure to property}
\begin{figure*}[htb]
\centering         
  \includegraphics[width=\linewidth]{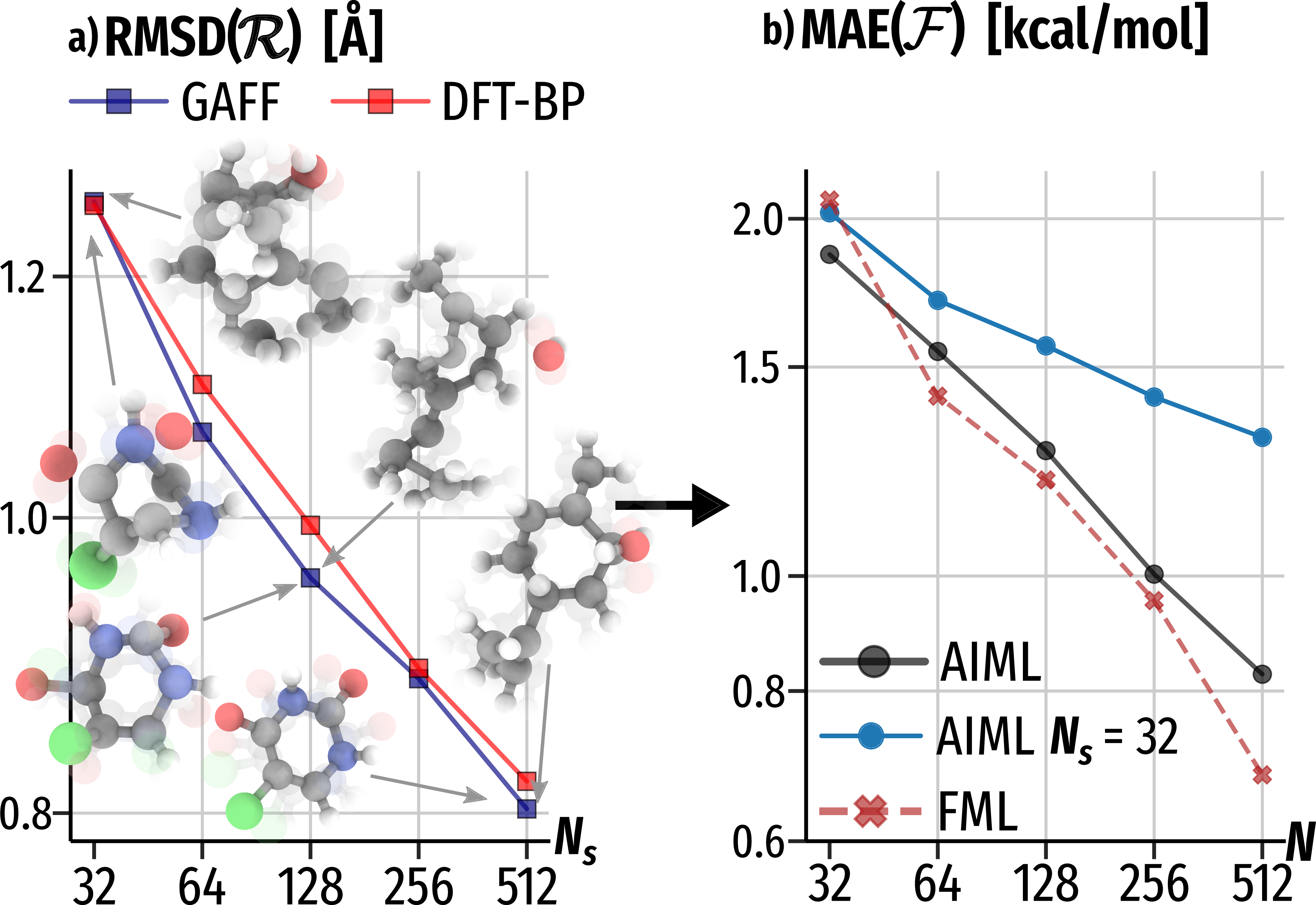}
  \caption{
  Learning curve of the Root-Mean-Square Deviation (RMSD), a measure of structural distance, of {\em ab initio} machine learning (AIML) three-dimensional structure predictions as a function of the number of training structures $N_s$ using GAFF2\cite{amber1, amber2} force field and density functional theory (DFT) with Becke-Perdew\cite{PhysRevA.38.3098, PhysRevB.33.8822} (BP) functional for training structure sampling (a). Two predicted average conformers $\mathbfcal{R}$ show the improvement of structure prediction along the learning curve (at $N_s=32,128,512$). The mean absolute error (MAE) of predicted free energies $\mathcal{F}$ of the FreeSolv\cite{FreeSolv} database as a function of $N$ for free energy machine learning (FML) (b) using MD sampling\cite{fml_paper} versus AIML (no sampling) and the Bag-of-Bonds\cite{bob} (BoB) representation. The AIML $N_{s}=32$ model was trained with only 32 structures for conformer prediction.}
  \label{fig:results}
\end{figure*}

In this section, we demonstrate the usefulness of AIML for the problem of accurate predictions of
free energies of solvation. 
In particular, we
focus on experimental free energies of solvation of 642 charge-neutral small to medium-sized bio-organic molecules, as encoded in the FreeSolv database~\cite{FreeSolv}. 
Solvation free energies\cite{gibbs,amber1, amber2, poissonB, pcm, TruhlarSMGB, smd,cosmo, cosmo2, KLAMT200043,doi:10.1021/jp971083h, KOVALENKO1998237, validationset} are of fundamental importance for chemistry, and the FreeSolv database has become a popular benchmark for performance
testing of novel models\cite{mlpcm,riniker2017molecular, axelrod2020molecular, doi:10.1021/acs.jcim.0c00479, hybridML, lim2020mlsolva,axelrod2020molecular,vermeire2020transfer, jinnouchi2019fly, venkatraman2018predicting}.

To use AIML to predict averaged structures and FML to predict free energies of solvation for out-of-sample molecules, we first trained AIML models as described above using molecular graphs as input and as labels for the averaged distances. 

These were obtained from extended force-field based MD runs and density functional theory (DFT) for conformer sampling with Def2TZVPD-FINE\cite{doi:10.1063/1.3484283, doi:10.1063/1.467146, Basisset1, Basisset2} basis set and Becke-Perdew\cite{PhysRevA.38.3098, PhysRevB.33.8822} (BP) functional in the gas-phase (s. sec.~\ref{sec:data} for details).
Note that we neglect the effect of water on the phase space of the solute (e.g. through hydrogen bonds). 
We believe that this aspect warrants further in-depth investigation within subsequent studies in the future.

Depending on the temperature (cfg. Fig.~\ref{fig:fig1}), some degrees of freedom do not get averaged out by the phase space integral (Eq.~\ref{eq:ensemble}), restricting the domain to certain local basins of the total free energy. These remaining degrees of freedom can be identified in the way described in Fig~\ref{fig:conformers}a). 
We demonstrate the idea of AIML for ambient temperatures in the gas phase for which conventional molecular graph topologies as in biochemistry hold -- without any loss of generality. 
Within such a regime, we can safely assume that any distance histogram matrix would be consistent with one topology which represents the coarse-grained back-bone that is not averaged out by the phase space integral. Of course, this approach could also be applied to any other set of conditions presuming that there is some way to easily infer valid topologies as a function of conditions (like temperature and composition). The latter is a separate problem that goes beyond the scope of this work.

As numerical evidence of the functionality of the AIML idea, we present 
in Fig. 3a) prediction errors of the three-dimensional Boltzmann averaged structures $ \mathbfcal{R}$ as a function of training set size (aka learning curves\cite{Vapnik1998, StatError_Muller1996,Cortes1993LearningCA}).

Numerical results shown in Fig.~\ref{fig:results} a) indicate a systematic linear improvement on a log-log scale\cite{vapnik1994learningcurves} as a function of the size of the training set $N_s$, that is, the number of averaged training structures. Note that $N$ stands for the number of training points for free energy values and that we have trained and evaluated two different machines after each other, for structure and free energy prediction respectively. For the maximal training set size considered ($N_{s}=512$ molecules, 80\% of FreeSolv), the average root-mean-square deviation\cite{jimmy, WALKER1991358,Kabsch:a12999} (RMSD), a measure of structural distance between structures, has decayed to only 0.80 {\AA} for predicted FF and 0.82 {\AA} for DFT averaged conformers, the slope of the learning curve, however, indicates that learning has not yet been saturated.
We find structure prediction of large molecules a particularly hard problem i.e. on average the RMSDs increase with the size of the structure. This was also observed for a random subset of the GDB17 molecular database \cite{gdb17} (s. SI. Fig. 7b for the scatter plot of molecular size vs. RMSD indicating a rough correlation).

Note that the corresponding learning curve predicting optimal (not Boltzmann-averaged) distances is less steep for the molecules in FreeSolv, and exhibits a higher offset (s. SI Fig.~4). This could be due to the fact that learning thermal averages is less ambiguous for AIML than learning potential energy minima as it is in the case of G2S. This also holds for the heavy atom-hydrogen distances (s. SI Tab.~1 and Fig. 1 in SI). From a different point of view, by computing the Boltzmann average over distances conformer flexibility is effectively integrated out, thus simplifying learning compared to the optimized structures.

Next, we demonstrate the learning efficiency (s. Fig.~\ref{fig:results}b) of AIML for free energy prediction based on previously predicted structures $\mathbfcal{R}$. The main disadvantage of the preceding free energy machine learning\cite{fml_paper} (FML) model was that it requires explicit conformer sampling for free energy prediction. The novel advantage of AIML is that no sampling is required. Instead, after prediction of the average conformer $\mathbfcal{R}$ the free energy prediction is based on the ML representation vector $\mathbf{x} (\mathbfcal{R})$ where $\mathbf{x}$ is the ML-based representation Bag-of-Bonds\cite{bob} (BoB).
Encouragingly, FML (requiring explicit MD sampling) and AIML exhibit similar learning curves, achieving mean absolute errors (MAE) of $\SI{0.68}{\kilo \cal \per \mol}$ and $\SI{0.82}{\kilo \cal \per \mol}$, respectively, as shown in Fig.~~\ref{fig:results}b). This indicates that the performance of distance-based representations in conjunction with AIML is fairly robust showing only $\SI{0.14}{\kilo \cal \per \mole}$ loss of accuracy compared to running MD simulations. Using predicted averaged DFT conformers we obtain roughly the same accuracy of $\SI{0.84}{\kilo \cal \per \mol}$ (s. SI. Fig. 8). This is consistent with our previous assessment of out-of-sampling AIML conformer RMSDs using FF and DFT which also resulted in comparable errors for both methods.
To illustrate the importance of structure prediction for subsequent property prediction, we have added a learning curve in Fig.~\ref{fig:results}b) using an AIML model with only $N=32$ average conformers for training the structure prediction. The mentioned model shows a much smaller learning rate. Generally, we find that better average structure prediction will lead to improved subsequent FML models (discussed in more detail in SI in Fig.~5).

We find AIML to perform consistently better when the training distances result from Boltzmann sampling instead of using optimized structures - similar to what we also noted above for the structure prediction (see SI Fig.~9). Even more surprising is the observation that training free energy prediction with AIML predicted structures resulted in slightly better models than training with the ground-truth averaged conformers resulting from MD. This indicates that AIML effectively smoothens conformer space by isolating the most important degrees of freedom, thus facilitating structure-based regression of thermal averages.

\subsection{Assessment of efficiency}

 \begin{figure}[htb]
          \centering         
          \includegraphics[width=\linewidth]{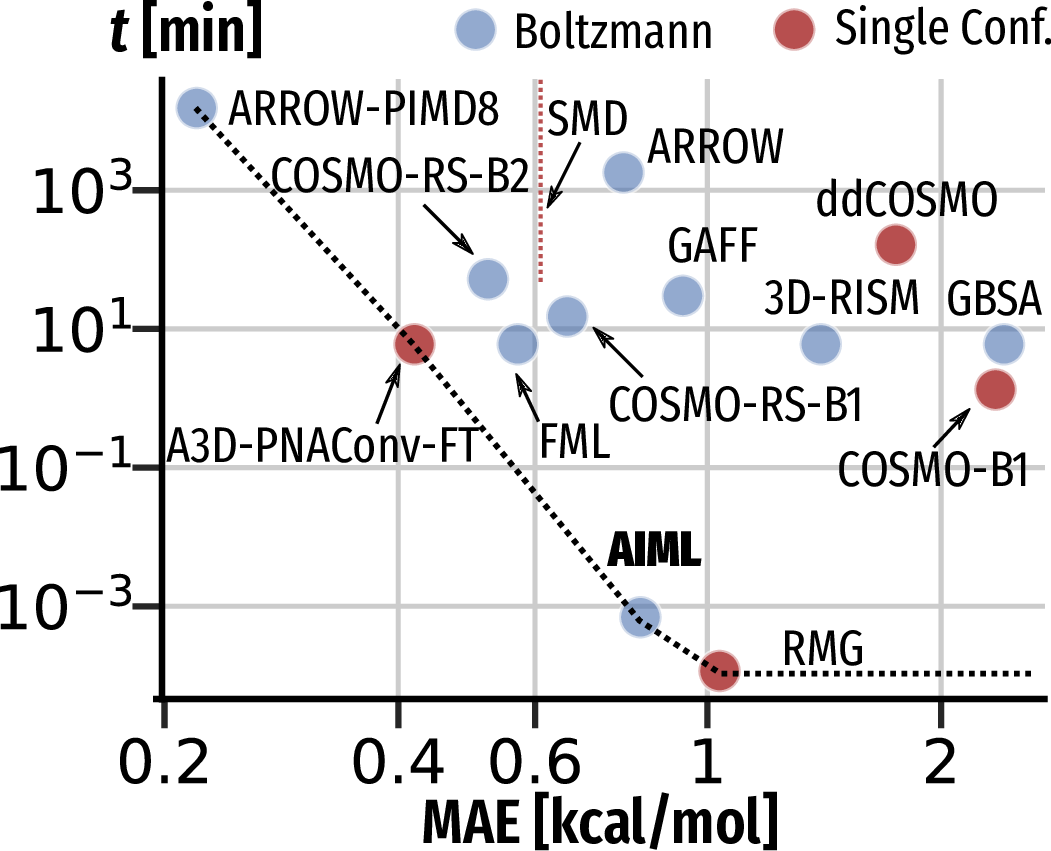}
          \caption{
                Comparison of solvation methods (if multiple weighted conformers used red else blue) in terms of mean absolute error (MAE) and order of magnitude of per molecule prediction time $t$ for FreeSolv\cite{FreeSolv} database. Pareto front (dotted) is formed by methods with best accuracy and cost per prediction trade-off. Central processing unit (CPU) compute time varies depending on hardware, code, etc., and was estimated if not available. All references and MAE of free energies are listed in the SI~Tab. 2.}
     \label{fig:g2fml_costcompare} 
 \end{figure}	

To gain a more comprehensive idea of AIML's value to the field, we have assessed the cost accuracy trade-off. 
Testing AIML on the FreeSolv\cite{FreeSolv} database we have measured average prediction times of $\SI{41}{\milli \second}$/molecule. 
These prediction times are dominated by the structure reconstruction task ($\SI{40}{\milli \second}$), 
while only $\SI{1}{\milli \second}$ is required to yield the free energy estimate (on a single-core AMD EPYC 7402P compute chip). For comparison, the corresponding prediction based on a classical force-field MD simulation protocol would have consumed three to four orders of magnitude more time, not to mention the costs associated with quantum chemistry based {\em ab initio} MD. 
To gain a comprehensive overview of the field, we have performed solvation free energy calculations for all of the FreeSolv molecules using the following methods (all listed free energies available, s. sec.~\ref{sec:data_avail}, MAEs listed in SI.~Tab. II):

\begin{enumerate}
    \item \underline{S}olvation \underline{m}odel based on \underline{d}ensity\cite{smd} (SMD) at M06-2X\cite{m06functional}/Def2-TZVPP\cite{doi:10.1063/1.3484283, doi:10.1063/1.467146, Basisset1, Basisset2} (timing for SMD implemented in Gaussian\cite{g16} not to be published)
    \item COSMO-RS-B1 and COSMO-RS-B2
    referring to COSMO-RS\cite{cosmo, cosmo2, cosmo3, cosmo4, cosmo5} with Def2TZVP\cite{doi:10.1063/1.3484283, doi:10.1063/1.467146, Basisset1, Basisset2} and
    Def2TZVPD-FINE\cite{doi:10.1063/1.3484283, doi:10.1063/1.467146, Basisset1, Basisset2} basis sets respectively and Becke-Perdew\cite{PhysRevA.38.3098, PhysRevB.33.8822} (BP) functional in the gas-phase
    \item \underline{R}eaction \underline{m}echanism \underline{g}enerator group (RMG) solvation\cite{chung_vermeire_wu_walker_abraham_green_2021}
    \item ddCOSMO\cite{ddcosmo} results obtained with PySCF\cite{https://doi.org/10.1002/wcms.1340} and PBE-0\cite{Perdew1996RationaleFM,doi:10.1063/1.478522}/Def2TZVPD\cite{doi:10.1063/1.3484283, doi:10.1063/1.467146, Basisset1, Basisset2}
    \item Generalized Born\cite{born, sasa} (GBSA) model, results obtained using AMBER\cite{amber1, amber2}
    \item Free energy machine learning\cite{fml_paper} (FML) with explicit conformer sampling on FreeSolv database 
\end{enumerate}
To complete the picture, we also included literature values for FreeSolv concerning the methods ARROW-PIMD8\cite{polarize},
\underline{T}hermodynamic \underline{i}ntegration (TI) with GAFF2\cite{amber1, amber2} extracted from the FreeSolv\cite{FreeSolv} database and \underline{r}eference \underline{i}nteraction \underline{s}ite \underline{m}odel\cite{3drism2,3drism3,validationset,3drism2,doi:10.1021/acs.jpca.0c06322} (3D-RISM).
The trade-off between cost and accuracy, including an outline of the resulting Pareto front, is displayed in Fig.~\ref{fig:g2fml_costcompare}.

We note that AIML adds to the convexity of the Pareto front, representing a meaningful compromise: Although roughly twice as slow, it is slightly more accurate at $\SI{0.82}{\kilo \cal \per \mole}$
than the Reaction Mechanism Generator\cite{rmg} (RMG) model (MAE of $\SI{0.98}{\kilo \cal \per \mole}$), 
but still four orders of magnitude faster than the 
next best \textit{ab initio} method COSMO-RS\cite{cosmo, cosmo2, cosmo3, cosmo4, cosmo5}. A list of all MAE is provided in SI Tab. 2. Thus, AIML is positioned on the Pareto front of the available solvation methods located in a sweet spot between speed and accuracy, providing the fastest predictions at the given accuracy of about $\SI{0.82}{\kilo \cal \per \mole}$. Note that the AIML learning curves have not yet saturated and that its accuracy will likely further improve if more training samples are included (c.f. ~Fig.~\ref{fig:results}).
Improving the AIML model will hardly worsen the prediction time due to the linear scaling of KRR predictions w.r.t. training set size (s.~sec.~\ref{sec:krr}) and therefore shift the Pareto front towards higher accuracy. Furthermore, it is important to note that arbitrary accurate \textit{ab initio} trajectories can be used for training while the prediction time is independent of the level of theory. We expect that recently published ML models tailored towards solvation such as SoluteML\cite{soluteML} may outperform the presented AIML models' accuracy, but we note that only a small training set of $N=512$ molecules was used and we expect MAE to decay further with the training set size. A3D-PNAConv-FT\cite{reviewer2c} combines the 2D and 3D structure and transfer learning achieving an MAE of $\SI{0.417}{\kilo \cal \per\mole}$ for the FreeSolv data set. Predictions require conformer sampling using a FF and the lowest-energy conformer. In contrast, AIML does not require sampling and is in fact replacing the functionality of a force-field or of an \textit{ab-initio} calculation.

Note that we have also tried to combine RMG and AIML/FML via the $\Delta$-ML approach\cite{deltaML} where RMG is used as a baseline for AIML, but unfortunately, the prediction errors did not improve (s. SI Fig.~8). Moreover, AIML performs worse for large molecules with many conformers (s. SI Fig. 7b). Unfortunately, combining a random sampled GDB17\cite{gdb17} dataset with 10000 molecules with the FreeSolv average conformers did not lead to improved structure predictions due to the small overlap of the two data sets (s. SI Fig.~6). Specifically, the small training set size and very high chemical diversity of the FreeSolv database including the elements C, H, O, S, N, F, I, Br, P, Cl, and up to 24 non-hydrogen atoms per molecule limit the accuracy of structure prediction for large compounds. Instead of adding random structures to improve structure prediction for the FreeSolv database (s. SI Fig.~6) we could show that sampling\cite{stoned} the local chemical space of the largest FreeSolv can help to improve the models' accuracy (s. SI Fig.~8). Alternatively, these problems may be resolved by improved graph-based representations that include information about local chemical substructures, leading to better structure and improved free energy predictions.

Recent graph-based ML models\cite{reviewer2a, reviewer2b} can achieve a competitive accuracy with root mean squared errors (RMSEs) around $\SI{1}{\kilo \cal \per \mole}$. We have achieved a similar RMSE of $\SI{1.35}{\kilo \cal \per\mol}$ for a training set size of $N=512$. We emphasize that the AIML approach is very different: First, AIML uses three-dimensional conformations which can lead to a much-improved accuracy (MAE of $\SI{0.57}{\kilo \cal \per \mole}$ for $N=490$) as we have shown earlier\cite{fml_paper} and allows going beyond fixed graphs. Secondly, AIML also predicts ensemble-based representation whereas SMILES-based ML use molecular graphs as input. We found a direct comparison with the two previously mentioned other ML methods\cite{reviewer2a, reviewer2b} difficult because they either use a different training-test split\cite{reviewer2a} or neglect\cite{reviewer2b} certain molecules of the FreeSolv database\cite{FreeSolv}. The comparison shows that our method might have a slightly higher initial offset due to having to learn the representation i.e. the averaged conformer before predicting the free energy. Our learning curves (s. Fig.~\ref{fig:results}) do not indicate saturation of the MAE with training set size $N$ and might still surpass graph-based models for large training set sizes because AIML contains information about molecular conformations.
    
\section{Conclusion}

\label{sec:conclusion}

We have introduced {\em ab initio} machine learning (AIML) allowing for efficient predictions of ensemble averages which systematically improve in accuracy as training set sizes grow. 
To the best of our knowledge, for the first time, AIML effectively bypasses the need for extensive MD or MC simulations to directly infer Boltzmann averaged geometries. 
Unlike all other solvation models (shown in Fig.~\ref{fig:g2fml_costcompare}) the AIML framework could easily be applied to other ensemble properties, e.g.~melting points, without much adaptation since no manual pre-selection of features for molecular fingerprints is required. 
AIML does not require any additional sampling for inferring ensemble averages of new out-of-sample query molecules: Instead AIML accounts for multiple Boltzmann weighted configurations implicitly through its training data. 
We have exemplified AIML for estimating experimental solvation free energies, and our numerical results amount to evidence showing that the conformer ensemble can effectively be linked to a single averaged conformer that serves as a canonical representative. 
AIML predictions are consistent with the previous free energy machine learning\cite{fml_paper} (FML) approach without the need to run an MD simulation for each prediction, reaching errors as low as $\SI{0.82}{\kilo \cal \per \mole}$ for 41 CPU-ms/molecule prediction cost.

Further analysis has revealed that AIML does not yet work well with all available molecular representations\cite{Parsaeifard_2021}. 
More specifically, we find that representations, tailored toward atomization energies and including explicit angular dependencies, such as FCHL19\cite{FCHL,felixFCHL}, yield less favorable AIML models (s. SI Fig.~9). 
Conversely, it might be possible to further improve AIML by tailoring and optimizing representations and architecture (e.g. using locality, symmetry, neural networks). 

The question of uniqueness is very fundamental for molecular representations\cite{PhysRevLett.109.059801,uniqueness_bing, uniqueness_Goedecker, uniqueness_Rupp, PhysRevLett.125.166001,fourier}. For the present dataset, however, this was not an issue as all averaged representations distinguished all data items.
It is possible, however, to imagine a scenario where this is not the case: For two different ensembles, $\mathcal{E}$ and $\bar{\mathcal{E} }$ with the same average conformer $\mathbfcal{R}= \bar{\mathbfcal{R}}$, but two different averages $\mathcal{A} \neq \bar{\mathcal{A}}$, the presented AIML model would make the same predictions and could not distinguish between these two systems. However, this problem could be resolved by including higher-order moments for the prediction of the representation, like including an AIML model for the standard deviation of the representation. Future work will deal with this question. Here our main focus was on molecules with uniquely defined bond topologies. For future applications, AIML could be applied to cases where assigning bond topologies is ambiguous or impossible such as transition states\cite{heini1, heini2} or molecules at very high temperatures. These are important cases where AIML can be used, but graph-based ML models cannot be used.

In summary, in comparison to classical or \textit{ab initio} MD-based predictions of free energies of solvation, AIML offers respective speed-ups by four to seven orders of magnitude. AIML achieves such speed-ups by effectively shifting computational cost for the query prediction to the training set generation. However, in light of the sheer scale of the chemical compound space available for molecular queries, this trade-off might be useful.

\subsection{Conformer and Free Energy Data}
\label{sec:data}

ML based on a single geometry can lead to ambiguous predictions\cite{fml_paper} ensemble property predictions because predictions can vary substantially depending on the conformer. Many body representations\cite{FCHL,felixFCHL, parsaeifard2021assessment, behler2011atom, Bircher_2021} rely on three-dimensional geometries, which becomes even more relevant if the target property depends on multiple relevant conformers. A solution to this issue is to sample configuration space to obtain a conformer invariant ML representations\cite{fml_paper}. Sampling can be achieved by different strategies: MD simulations\cite{amber1, amber2}, systematic conformer ensemble scans\cite{crest1, crest2, crest3} and conformer generation methods either knowledge or force field (FF) based such as ETKDG6\cite{Yoshikawa2019FastEF}, Gen3D \cite{doi:10.1021/acs.jcim.5b00654} and others\cite{Miteva2010Frog2E3,doi:10.1021/ci100031x,landrum2006rdkit,confab,ballon1}.
A more expensive method but accurate method is to obtain conformations using \textit{ab initio} approaches such as density functional theory (DFT)\cite{kresse1,kresse2} or tight binding\cite{crest1, crest2, crest3} (TB). Despite these advantages, a common pitfall of these methods is that sometimes extension to arbitrary chemistries is not straightforward. To this end, ML-based methods\cite{lemm2021machine, Mansimov2019MolecularGP, Kstner2009DLFINDAO, bayesian} hold the promise of providing faster and more general structure predictions. There are only very few ML structure generation methods e.g., based on reinforcement learning\cite{meldgaard2021generating} or stochastic normalizing flows\cite{snf} that take energy weights of different conformers into account. Here, to obtain a diverse set of conformers as the AIML training set we have performed MD simulations in vacuum at an elevated temperature of $T=\SI{350}{\kelvin}$ using OpenMM\cite{openmm} using a Langevin integrator. GAFF2\cite{amber1, amber2} with a time-step of $\Delta t = \SI{2}{\femto \second }$ was used with a total simulation time of $\SI{2}{\nano \second}$. Partial charges are computed with antechamber\cite{amber1,amber2} at AM1-BCC\cite{bcc} level. MD samples are selected with $\SI{2}{\pico \second}$ time separation. To compare AIML with COSMO-RS\cite{cosmo, cosmo2, cosmo3, cosmo4, cosmo5} solvation method, we used the COSMO-RS workflow based on \textit{ab initio} DFT calculations with Turbomole\cite{TURBOMOLE} and the Becke-Perdew (BP)\cite{PhysRevA.38.3098, PhysRevB.33.8822} functional as implemented in COSMOconf\cite{COSMOconf} with two different basis sets, Def2TZVP and Def2TZVPD-FINE\cite{doi:10.1063/1.3484283, doi:10.1063/1.467146, Basisset1, Basisset2} (for future reference referred to as B1 and B2). Based on these results, free energies are extracted using the COSMOtherm\cite{therm} program. In addition, the \underline{r}eaction \underline{m}echanism \underline{g}enerator group (RMG) based approach was used to compute free energies of solvation\cite{chung_vermeire_wu_walker_abraham_green_2021} of the FreeSolv database via the \texttt{leruli.com} API\cite{leruli}. The FreeSolv\cite{FreeSolv} dataset contains 642 charge neutral compounds and their experimental free energies of solvation. The average unsigned error of the experimental values is $\SI{0.57}{\kilo \cal \per \mol}$ close to the level of thermal energy fluctuations ($k_{B} \cdot \SI{300}{\kelvin} \approx \SI{0.6}{\kilo \cal \per \mol} $). All ML models use a maximal training set size of $80\%$ corresponding to $N = 512$ molecules. Hyperparameters are optimized with nested five-fold cross validation.

\section{Data and Code availability}
\label{sec:data_avail}
The AIML code and all free energies of solvation of the FreeSolv database (if produced by the authors) are published in a freely available repository \url{https://doi.org/10.5281/zenodo.6401711}. We gladly provide more data for specific requests.

\section{Acknowledgements}
We acknowledge support from the European Research Council (ERC-CoG Grant QML). This project has received funding from the European Union's Horizon 2020 research and innovation program under Grant Agreement \#772834. 
This research was also supported by the NCCR MARVEL,
a National Centre of Competence in Research, funded
by the Swiss National Science Foundation (grant number
182892).

\section{Declaration of Conflicting Interests}
DL, GFvR, and OAvL are shareholders of Leruli GmbH.


\bibliographystyle{apsrev4-1} 
\bibliography{main} 

\end{document}